\PassOptionsToPackage{square,sort&compress,comma,numbers}{natbib}
\documentclass[final,5p,times,twocolumn]{elsarticle}
\usepackage{amssymb}
\usepackage{lipsum}
\usepackage{amsmath}
\journal{Physics Letters B}

\begin{document}

\begin{frontmatter}

%% Title, authors and addresses

%% use the tnoteref command within \title for footnotes;
%% use the tnotetext command for theassociated footnote;
%% use the fnref command within \author or \affiliation for footnotes;
%% use the fntext command for theassociated footnote;
%% use the corref command within \author for corresponding author footnotes;
%% use the cortext command for theassociated footnote;
%% use the ead command for the email address,
%% and the form \ead[url] for the home page:
%% \title{Title\tnoteref{label1}}
%% \tnotetext[label1]{}
%% \author{Name\corref{cor1}\fnref{label2}}
%% \ead{email address}
%% \ead[url]{home page}
%% \fntext[label2]{}
%% \cortext[cor1]{}
%% \affiliation{organization={},
%%            addressline={}, 
%%            city={},
%%            postcode={}, 
%%            state={},
%%            country={}}
%% \fntext[label3]{}

\title{Thermodynamics of EMD-AdS black hole with deformed horizon}

%% use optional labels to link authors explicitly to addresses:
%% \author[label1,label2]{}
%% \affiliation[label1]{organization={},
%%             addressline={},
%%             city={},
%%             postcode={},
%%             state={},
%%             country={}}
%%
%% \affiliation[label2]{organization={},
%%             addressline={},
%%             city={},
%%             postcode={},
%%             state={},
%%             country={}}

\author[1]{Yusheng Z. He}
\ead{2022022181@m.scnu.edu.cn}

\author[1,2]{Jia-Hui Huang \corref{cor1}}
\cortext[cor1]{Corresponding author.}
\ead{huangjh@m.scnu.edu.cn}

\affiliation[1]{organization={Department of Physics},%Department and Organization
            addressline={South China Normal University},
            city={Guangzhou},
            postcode={510006},
            country={China}}

\affiliation[2]{organization={Guangdong Provincial Key Laboratory of Quantum Engineering and Quantum Materials,  Guangdong-Hong Kong Joint Laboratory of Quantum Matter},%Department and Organization
            addressline={South China Normal University}, 
            city={Guangzhou},
            postcode={510006}, 
            country={China}}

\author[3]{Jinbo Yang }
\ead{yangjinbo@gzhu.edu.cn}

\affiliation[3]{organization={Department of Astronomy, School of Physics and Materials Science},%Department and Organization
            addressline={Guangzhou University}, 
            city={Guangzhou},
            postcode={510006}, 
            country={China}}

\begin{abstract}

We show that the horizon shapes of static Einstein-Maxwell-dilaton (EMD) black holes can be deformed through an approach analogous to those observed in novel topological black hole solutions supported by two massless axion fields. Particularly, the radial part of the spacetime remains unchanged, while the angular part has a non-constant Ricci scalar, signaling the presence of anisotropy. We further explore the implications of this anisotropy for the thermodynamic properties of the black hole. The validity of the Misner-Sharp (MS) mass and the unified first law are also addressed in this study.
\end{abstract}

%%Graphical abstract
%\begin{graphicalabstract}
%\includegraphics{grabs}
%\end{graphicalabstract}

%%Research highlights
%\begin{highlights}
%\item Research highlight 1
%\item Research highlight 2
%\end{highlights}

\end{frontmatter}

%\tableofcontents

%% \linenumbers

%% main text

\section{Introduction}
\label{introduction}
%Since general relativity and quantum field theory build, they have played a crucial role in our understanding of physical phenomena. However, their most significant products, black hole physics and gauge field theory, appear to be fundamentally distinct. Finding a unified framework that perfectly integrates these two theories remains one of the most important unresolved issues in theoretical physics. In 1998, the holographic duality emerged, pointing out the equivalence between the Anti-de Sitter (AdS) black hole theory and the Conformal Field Theory (CFT). This has perhaps paved a way for exploring quantum gravity.%
The AdS/CFT correspondence provides a novel approach to studying strongly coupled gauge theories by transforming the strong coupling problems into weakly coupled gravitational problems \citep{Maldacena_1999,Gubser:1998bc,Witten:1998qj}. AdS black holes and their thermodynamics have thus become one of the most studied black hole theories in recent years (for a review, see \cite{Hubeny:2014bla, huang2024adscftdualityholographicrenormalization}). 

%This is particularly relevant in the context of AdS/CFT duality, where the gravitational dynamics in the bulk are dual to the field theory dynamics on the boundary.%
\par
One class of AdS black holes are associated with EMD systems. These systems, which couple a Maxwell field and a dilaton field, have received widespread attention due to their ability to holographically describe systems with finite charge density or condensed matter \cite{Chan:1995,Cai:1996,Cai:1998,Charmousis:2002,Gouteraux:2011qh,Fu:2020oep,Bai:2024, Priyadarshinee:2021rch,Mahapatra:2020wym}. 
For instance, the holographic description of these black holes has been used to explore the properties of strongly coupled field theories at finite temperature and density.

\par
A notable class of EMD-AdS black holes is found in STU-like gauged supergravity models. A variety of analytical solutions in D dimensions have been found for this class of solutions \cite{chow2011singlerotationtwochargeblackholes,Wu:2011zzh,Liu:2012ed,Lu:2013eoa}. In the study of AdS black holes within STU-like models, a particular analytical solution has been repeatedly identified with varying ansatz and parameters \cite{Nozawa:2020gzz,Anabalon:2013sra,Feng:2013tza,Ren:2019lgw,rao2024hairyblackholesarbitrary}. A form of the scalar potential is given by \cite{Feng:2013tza}
%(改这一段，三个都提（冯，任，饶）)%
\begin{equation}
\begin{split}
    \mathcal{V}=&-(g^{2}-\alpha)e^{-\frac{1-\mu}{\nu}\phi}\bigg[(\mu-1)(2\mu-1)e^{\frac{2}{\nu}\phi}-4(\mu^{2}-1)e^{\frac{1}{\nu}\phi}\\
    &+(\mu+1)(2\mu+1)\bigg]-\alpha e^{-\frac{1+\mu}{\nu}\phi}\bigg[(\mu+1)(2\mu+1)e^{\frac{2}{\nu}\phi}\\
    &-4(\mu^{2}-1)e^{\frac{1}{\nu}\phi}+(\mu-1)(2\mu-1)\bigg]\,.
\end{split}
\end{equation}
The scalar potential employed in this paper is also a specific branch of it. %写一句过程见附录A%
\par
On the other hand, AdS background permits a black hole with topologically non-spherical horizon \cite{Mann:1997,Birmingham:1998,Emparan:1999,Birmingham:2007}. %unlike maximally symmetric spacetimes \cite{Hawking:1972}such as AdS spacetime,%
Furthermore, the maximal symmetry can be broken by introducing holographic axions \cite{Andrade:2013gsa, Donos:2014, Arias:2017, Jiang:2017, Baggioli:2019, Liu:2023}. Some of these solutions constructed by the method proposed in Refs.\cite{Zhang:2013tca,Yang:2023nnk} allow the Ricci scalar of the angular space $\hat{R}$ to be non-constant, instead depending on angular coordinate. 
%Similar construct can be found in \cite{Peng:2021xwh, Peng:2022ttn,Peng:2024sbr}. %

The thermodynamics of topological black holes has also been extensively studied \cite{Priyadarshinee:2021rch, Tian:2014, Tian:2019, Andrade:2013gsa, Wang:2022err,Kong:2022tgt,Mancilla:2024spp}. We will take one of these solutions as an example to demonstrate that the angular part of such deformed solutions can be combined with the radial part of EMD black holes through a warped product, forming a new metric that satisfies the Einstein equations. The interesting aspect of these new solutions is that the deformation does not alter the action and thermodynamic quantities of the EMD black holes, but only changes the curvature of the angular space. Referring to the unified first law \cite{Hayward:1993wb,Hayward:1998,Cai:2005}, we introduce a quasi-local method to study the thermodynamics of the deformed solutions. This approach is similar to the one used in the context of perfect fluid dark matter black holes in a phantom background, where the MS mass \cite{Misner:1964, Hayward:1996} is identified as the total energy of the spacetime within a certain radius, and it helps in formulating the corresponding thermodynamic first law.

Our paper is organized as follows. In Section \ref{sec.2}, we present the analytical solution of the EMD-AdS hyperbolic black hole with axions. In Section \ref{sec.3}, taking temperature as an example, we study the changes in the thermodynamic quantities of the black hole due to the angular deformation. In Section \ref{sec.4}, we introduce the global and quasi-local viewpoint to derive the first law of thermodynamics for the deformed black hole and discuss the mass parameter. Finally, we give a summary and discuss some open questions.

\section{Einstein-Maxwell-Dilaton Black Hole with Axions}
\label{sec.2}

We consider the following Lagrangian:

\begin{equation}
\begin{split}
\mathcal{L}=&\frac{R}{16 \pi G}-\frac{1}{16} e^{-\alpha  \phi }F^{\mu  \nu }F_{\mu  \nu }\\&-\frac{1}{2G}\nabla_{\mu }\phi \nabla ^{\mu }\phi-\frac{1}{G}\mathcal{V}(\phi)-\frac{1}{2}\sum_{i=1}^2 \nabla _{\mu }\psi ^i\nabla ^{\mu }\psi ^i\,,
\end{split}
\end{equation}
where $F_{\mu \nu }=(\text{dA})_{\mu \nu }$ describes the strength of the U(1) gauge field $A_{\mu}$, ${\phi}$ is the dilaton field, and ${\psi}^{i}$ with $i=1,2$ are two massless scalars in angular space. The potential of the dilaton field is:
\begin{equation}
\mathcal{V}(\phi)=-\frac{2 \left(8 \alpha ^2 e^{\frac{1}{2} \left(\alpha -\frac{1}{\alpha }\right) \phi}+\left(3 \alpha ^2-1\right) \alpha ^2 e^{-\frac{\phi}{\alpha }}+\left(3-\alpha ^2\right) e^{\alpha  \phi}\right)}{\left(\alpha ^2+1\right)^2 L^2}\,,
 \end{equation} 
where ${\alpha}$ is a real parameter and $L$ can be read as cosmological constant ${\Lambda}$. Some special values of ${\alpha}$ are correspond to totally different solutions. For example, ${\alpha}=0,\frac{1}{\sqrt{3}}, 1$, and $\sqrt{3}$ correspond to special cases of STU supergravity. This potential was found in \cite{Ren:2019lgw} and it is proportional to the $U({-\phi})$ in \cite{Feng:2013tza} or \cite{rao2024hairyblackholesarbitrary}. 

\par
We perform variation on the action with respect to the metric $g_{\mu\nu}$, the U(1) electromagnetic field $A_{\mu}$, and the dilaton field $\phi$ respectively. The equations of motion are:

 \begin{equation}
   G_{\mu \nu}=R_{\mu \nu }-\frac{1}{2}{R g_{\mu \nu }}=8 \pi G\left({T^{(em)}}_{\mu \nu}+{T^{(\phi)}}_{\mu \nu }+{T^{(\psi^{i})} }_{\mu \nu }\right)\,,
   \label{eqn:ein eqn}
 \end{equation} 
 \begin{equation}
\nabla ^{\mu }(e^{-\alpha\phi}F_{\mu \nu })=0\,,
   \label{eqn:max eqn}
 \end{equation} 
  \begin{equation}
\nabla ^{\mu }\nabla _{\mu }\phi=\frac{\partial {\mathcal{V}}}{\partial \phi }\,,
   \label{eqn:KG eqn dilaton}
 \end{equation}
  \begin{equation}
\nabla ^{\mu }\nabla _{\mu }{\psi}^{i} =0\,,
   \label{eqn:KG eqn axion}
 \end{equation}
where

  \begin{equation}
{T^{(em)}}_{\mu \nu }={e^{-\alpha  \phi} \left(F_{\mu \sigma } {F_{\nu }}^{\sigma }-\frac{1}{4} F_{\rho \sigma } F^{\rho \sigma } g_{\mu \nu }\right)}\,,
 \end{equation} 
 \begin{equation}
{T^{(\phi)}}_{\mu \nu }=\frac{1}{G}[{\partial_{\mu}} {\phi }{\partial_\nu } { \phi }-g_{\mu \nu }\left(\frac{1}{2}{\partial_\alpha \phi }\partial ^{\alpha }\phi +\mathcal{V} (\phi) \right)]\,,
 \end{equation} 
 \begin{equation}
 {T^{({\psi}^i)}}_{\mu \nu }=({\partial_{\mu}} {{\psi}^i }{\partial_\nu } { {\psi}^i }-\frac{1}{2}g_{\mu \nu }{\partial_\alpha {\psi}^i }\partial ^{\alpha }{\psi}^i )\,.
\end{equation} 
Equation (\ref{eqn:ein eqn}) is the Einstein equation with all 4 energy-momentum tensors. Equation (\ref{eqn:max eqn}) is the Maxwell equation. (\ref{eqn:KG eqn dilaton}) and (\ref{eqn:KG eqn axion}) are Klein-Gordon(KG) equations for dilaton and axions. 
We will discuss the solutions of these equations in detail.
\par

The ansatz we take to solve equations of motion is:

\begin{equation}
\text{ds}^2=-f(r)\text{dt}^2 +\frac{\text{dr}^2}{f(r)}+U(r) \text{d$\Sigma $}_{2,\,k}^2\,,
\label{eq:ansatz}
 \end{equation}
where
 \begin{equation}
\text{d$\Sigma $}_{2,k}^2=\frac{\text{d$\rho $}^2}{1-k \rho ^2-e(\rho )}+\rho ^2\text{d$\varphi $}^2\,.
\label{eq:ansatz angular}
\end{equation} 
The $e(\rho)$ term is from Yang's solution I in \cite{Yang:2023nnk}. We use $\text{d$\Sigma $}_{2,k}^2$ to denote two-dimensional angular space where $k=1, 0, -1$ correspond to different curvatures of the $\Sigma$. Properties of these cases in Anti-de Sitter spacetime have been discussed in \cite{Witten:1998qj, Hawking:1983, Birmingham:1998}.

\par
There are 7 functions ${f(r), \phi(r), U(r), A_{t} (r),e (\rho), {\psi}^{1} (\rho),  {\psi}^{2} (\varphi)}$ to be solved. The KG equation of ${\psi}^{2}$ is

\begin{equation}
\frac{d^2 {\psi}^2(\varphi)}{d{\varphi}^2}=0\,,
\end{equation}
and it can be solved simply by $ {\psi}^{2} (\varphi)=\xi \varphi$. We get five ODEs from Einstein-Maxwell equations and another one from KG:
 \begin{equation}
 \begin{split}
   V(\phi)&+8\pi G e^{-\alpha  \phi} {A_t}'^2+\frac{f' U'+f U''-2 \text{k}}{U}\\
           &+\frac{1}{2U}(-\frac{2}{\rho}\frac{de}{d{\rho}}+16\pi G(1-k \rho ^2-e)(\frac{d{\psi}^1}{d{\rho}})^2+\frac{16\pi G}{{\rho^2}}(\frac{d{\psi}^2}{d{\varphi}})^2)=0\,,
 \end{split}
 \label{eq:ein eqn fenliang}
 \end{equation} 
 \begin{equation}
16\pi G e^{-\alpha  \phi } {A_t'}^2-f''+\frac{f U''-2 k}{U}=0\,,
 \end{equation} 
  \begin{equation}
U'^2-U \left(2 U''+U \phi '^2\right)=0\,,
 \end{equation}
 
   \begin{equation}
(1-k \rho ^2-e)(\frac{d{\psi}^1}{d{\rho}})^2-\frac{1}{{\rho^2}}(\frac{d{\psi}^2}{d{\varphi}})^2=0\,,
\label{eq:EOM of axion}
 \end{equation}
 
  \begin{equation}
-\alpha  \phi ' A_t'+\frac{U'}{U} A_t'+A_t''=0\,,
 \end{equation}
 
   \begin{equation}
\left(\rho  \frac{de}{d{\rho}}+2 e+4 \text{k} \rho ^2-2\right)\frac{d{\psi}^1}{d{\rho}}-2 \rho  \left(1-k \rho^2-e\right)\frac{d^2{\psi}^1}{d{\rho}^2}=0\,,
\label{eq:EOM of angular}
 \end{equation}
where we use $'$ to denote the derivative of the function with respect to $r$. Equation (\ref{eq:ein eqn fenliang}) can be easily separated into radial and angular parts. So we can set $(\ref{eq:ein eqn fenliang})_{rad}=-(\ref{eq:ein eqn fenliang})_{ang}=\eta$. It means that Einstein equations (\ref{eqn:ein eqn}) should also hold when $e(\rho)={\psi}^i=0$. So we get a new equation from angular parts of equations of motion:
\begin{equation}
  \begin{cases}
     -\frac{2}{\rho}\frac{de}{d{\rho}}+16\pi G(1-k \rho ^2-e)(\frac{d{\psi}^1}{d{\rho}})^2+\frac{16\pi G}{{\rho^2}}(\frac{d{\psi}^2}{d{\varphi}})^2=-\eta\\
     \mathcal{V}(\phi)U+8\pi G e^{-\alpha  \phi} {A_t}'^2 U+f' U'+f U''-2 \text{k}=2\eta
  \end{cases}
  \,.
  \label{eq:new EOM of axion}
\end{equation}
Comparing equation (\ref{eq:EOM of axion}) and (\ref{eq:new EOM of axion}), we can solve them for $e(\rho)$ and ${\psi}^1 (\rho)$:
\begin{equation}
  \begin{cases}
    e(\rho)=16\pi G {\xi}^2 \log(\rho)-\eta \rho^2+Const.\\
    \frac{d{\psi}^1}{d{\rho}}=\frac{\xi}{\rho \sqrt{1-k {\rho}^2-e}}
  \end{cases}
  \,.
\end{equation}
We can consider the constant in $e(\rho)$ as $e(\rho)=16\pi G{\xi}^2 \log{(\zeta \rho)}-\eta \rho^2$. These solutions can satisfy equation (\ref{eq:EOM of angular}) automatically. The analytic solutions of ${f(r), \phi(r), U(r), A_{t} (r)}$ are:
 \begin{equation}
f=\left(\text{k}+\eta-\frac{16 \pi Gc}{r}\right)\left(1-\frac{b}{r}\right)^{\frac{1-\alpha ^2}{1+\alpha ^2}}+\frac{r^2 }{L^2} \left(1-\frac{b}{r}\right)^{\frac{2 \alpha ^2}{\alpha ^2+1}}\,,
\label{eq:f}
 \end{equation} 
 
  \begin{equation}
U=r^2 \left(1-\frac{b}{r}\right)^{\frac{2 \alpha ^2}{\alpha ^2+1}}\,,
\label{eq:U}
 \end{equation}
 
   \begin{equation}
A_t=2\sqrt{\frac{b c}{\alpha ^2+1}} \left(\frac{1}{r_h}-\frac{1}{r}\right)\,,
\label{eq:At}
 \end{equation}
 
  \begin{equation}
\phi =\frac{2 \alpha}{\alpha ^2+1} \ln \left(1-\frac{b}{r}\right)\,.
\label{eq:phi}
 \end{equation}
Thus, the nontrivial $\eta$ in the function can be simplified as $k=k+\eta$ so it can be seen as shift the $k$ in metric. The solution has four parameters b(dilaton), c(charge) and $\xi$, $\zeta$(axions) in addition to $\alpha$ and $k$. This system is invariant under $\alpha\rightarrow-\alpha$ and $\phi \rightarrow -\phi$. We will discuss them in radial and angular parts separately.
\par
In the angular part, its independent line element ${d\Sigma }_{2,k}^2$ seems to be described by three parameters $k, \xi, \zeta$, but we can rescale angular space by setting $\zeta$ to 1. Therefore, the Ricci scalar of angular space is:
\begin{equation}
    \hat{R} (\rho)=2k+16\pi G\frac{\xi^2}{\rho^2}\,,
    \label{angular R}
\end{equation}
which is a function of $\rho$ rather than a constant, and we use a hat to denote the angular part of spacetime. This Ricci scalar indicates that its geometry is determined by two parameters $k$ and $\xi$ and there is probably a singularity at $\rho =0$. As we know, $\hat{g}_{\rho \rho}$ should be positive to ensure the correct signature of the metric. This shows:
\begin{equation}
    k< \frac{1-16\pi G\xi^2\ln \rho}{\rho^2}\,.
\end{equation}
The above inequality can be taken as $k(\rho)$, a function of $\rho$. It determines the minimum value of $|k|$:
\begin{equation}
    |k|_{\text{min}}=8\pi G \xi ^2 e^{-1-\frac{1}{8\pi G\xi ^2}}\,,
\end{equation}
at the location
\begin{equation}
    \rho=e^{\frac{1}{2}+\frac{1}{4\pi G\xi ^2}}\,,
\end{equation}
by setting $k'(\rho)=0$. In other words, the roots of $\frac{1}{\hat{g}_{\rho \rho}}=0$ are singularities of angular space and they determine the range of $\rho$.

\par
In $t-r$ part, we will focus on radial functions in (\ref{eq:f})-(\ref{eq:phi}). If we take $b=0$, $\phi$ and $A_t$ will be trivial and the solution will become Schwarzschild-AdS solution with axions:
\begin{equation}
 \begin{split}
     ds^2=-(k-&\frac{16\pi G c}{r}+\frac{r^2}{L^2})dt^2+(k-\frac{16\pi G c}{r}+\frac{r^2}{L^2})^{-1} dr^2\\
     &+r^2 (\frac{d \rho^2}{1-k \rho^2-e(\rho)}+\rho^2 d\varphi^2)\,.
 \end{split}
\end{equation}
If we take c=0, the electric field will disappear, but the dilaton field $\phi$ will still be nontrivial. This solution is an uncharged black hole with scalar hair. In this case, $f$ becomes:
\begin{equation}
    f=k \left(1-\frac{b}{r}\right)^{\frac{1-\alpha ^2}{1+\alpha ^2}}+\frac{r^2 }{L^2} \left(1-\frac{b}{r}\right)^{\frac{2 \alpha ^2}{\alpha ^2+1}}\,.
\end{equation}
If $k>0$, it will no longer describe a black hole. So we focus on the case of $k<0$. We will calculate thermodynamic quantities of the black holes with the $k \neq -1$ with dilaton field in the following. We set the AdS radius $L=1$.

\par
Interestingly, the deformation with axions can be applied to any spherical, planar or hyperbolic solution, not only to the RN or EMD system. Deforming the black hole horizon might have wide application in AdS/CFT. We will study the angular deformation of rotating black hole in our future work.

\section{Horizon Temperature}
\label{sec.3}

Consider the situation where c=0 and $k<0$ first. The analytic solution of EMD deformed black hole is given by the ansatz in (\ref{eq:ansatz}) and (\ref{eq:ansatz angular}). The two curvature singularities of r are at $r=0$ and $r=b$, and the horizon of the black hole is determined by $f(r_h)=0$. From horizon condition, we get the parameter $b$ as a function of $k$ and $rh$ :
\begin{equation}
    b=r_h-(-k)^{\frac{\alpha ^2+1}{3 \alpha ^2-1}} r_h^{\frac{3-\alpha ^2}{1-3 \alpha ^2}}\,.
    \label{eq:b}
\end{equation}
Equation (\ref{eq:b}) shows that discussing the case of $k<0$ is simpler than spherical-like situation and it will reduce to results in \cite{Ren:2019lgw} when $k=-1$. From Section \ref{sec.2}, we know that axions do not change physics of the t-r part of the black hole. But event horizon is special: Shape of horizon is determined by ${d\Sigma }_{2,k}^2$ and its position $r_h$ is determined by $f(r)$. In fact, deformation of angular space changes the meaning of r. For example, in $\mathbb{R}^2$, spherical $r^2=x^2+y^2$ and hyperbolic $r^2=x^2-y^2$. A viewer from infinite distance can assume that the black hole is wrapped in layers of hypersurfaces of $r=r_a (r_h<r_a<\infty)$. Axions just deform the shape of hypersurface and the specific deformation of horizon's shape has been discussed in detail. The thermodynamic discussion with angular deformation is meaningful because only some of its properties have changed. We will study the temperature of horizon first. It is given by
\begin{equation}
\begin{split}
        T&=\frac{f'(r_h)}{4\pi}\\
         &=\frac{1}{4\pi(1+ \alpha ^2)}[\left(3-\alpha ^2\right) (-k)^{\frac{2 \alpha ^2}{3 \alpha ^2-1}} r_h^{\frac{1+\alpha ^2}{1-3 \alpha ^2}}\\
         &+\left(3 \alpha ^2-1\right) (-k)^{\frac{\alpha ^2-1}{3 \alpha ^2-1}} r_h^{-\frac{1+\alpha ^2}{1-3 \alpha ^2}}]
         \label{eq:T}\,,
\end{split}
\end{equation}
where $(-k)$ appears in temperature with the exponent $\alpha$. From exponent in equation (\ref{eq:T}), there are two special $\alpha$: $\sqrt{3}$ and $1/\sqrt{3}$. In the case $\alpha=\frac{1}{\sqrt{3}}$, equation (\ref{eq:T}) is no longer valid and $f$ will become:
\begin{equation}
    f_{\alpha=1/\sqrt{3}}=\sqrt{1-\frac{b}{r}} \left(k+r^2\right)\,,
\end{equation}
so that there is a real root of $f$ when $k<0$: $r_h=\sqrt{-k}$, and the temperature is $T=\frac{\sqrt{(-k)-b\sqrt{-k}}}{2\pi}$. It will reach zero when $b=\sqrt{-k}$. When $\alpha=\sqrt{3}$, f will become:
\begin{equation}
    f_{\alpha=\sqrt{3}}={(1-\frac{b}{r})}^{-\frac{1}{2}} \left[k+(r-b)^2\right]\,,
\end{equation}
\begin{figure}[!t]
    \centering
    \begin{minipage}[c]{0.4\textwidth}
        \centering
        \includegraphics[scale=0.64]{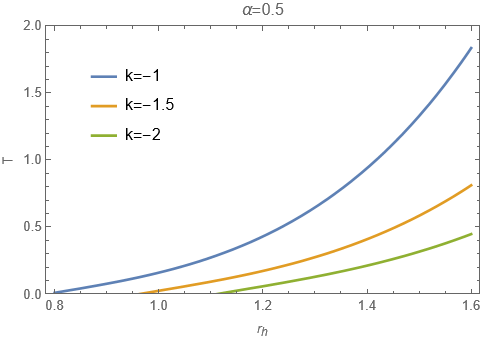}
    \end{minipage}
    \hspace{0.3\textwidth}
    \begin{minipage}[c]{0.4\textwidth}
        \centering
        \includegraphics[scale=0.64]{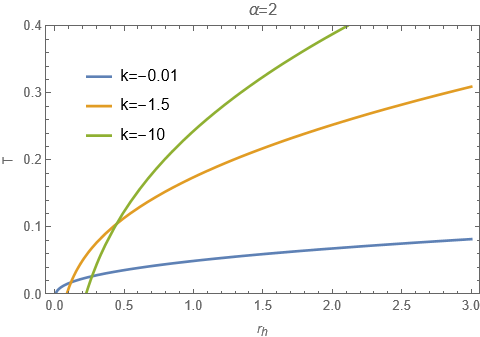}
    \end{minipage}
    \caption{There is one branch of black hole solution. In this case, T is a monotonically increasing function of $r_h$ and it can reach zero. In the top plot($\alpha<1/\sqrt{3}$), at the same T, larger $|k|$ leads to larger $r_h$. In the bottom plot($\alpha > \sqrt{3})$, the curves of different k will have a intersection. For negative $k$, the $r_h$ of $T=0$ will always be positive.}    \label{fig.1}
\end{figure}
where $b=r_h-\sqrt{-k}$ and $T=\frac{(-k)^{\frac{1}{4}}}{2\pi} \sqrt{r_h}$. The temperature is a power function of $r_h$. Except for these two exceptions, temperature can be a function of $r_h$ with parameter $K$. The effect of different k on horizon temperature under $\alpha < 1/\sqrt{3}$ and $\alpha > \sqrt{3}$ is shown in Fig.\ref{fig.1}. In this case, the temperature reaches zero when 
\begin{equation}
    r_h=(-k)^{\frac{1}{2}} \left(\frac{1-3 \alpha ^2}{3-\alpha ^2}\right)^{\frac{1-3 \alpha ^2}{2 \left(\alpha ^2+1\right)}}\,.
\end{equation}
The extremal black hole with scalar hair is given by c=0 and
\begin{equation}
    b=\sqrt{-k} \frac{2 \left(\alpha ^2+1\right) }{3-\alpha ^2}\left(\frac{3 \alpha ^2-1}{\alpha ^2-3}\right)^{\frac{1-3 \alpha ^2}{2 \alpha ^2+2}}\,.
\end{equation}
\par
\begin{figure}
    \centering
    \includegraphics[scale=0.64]{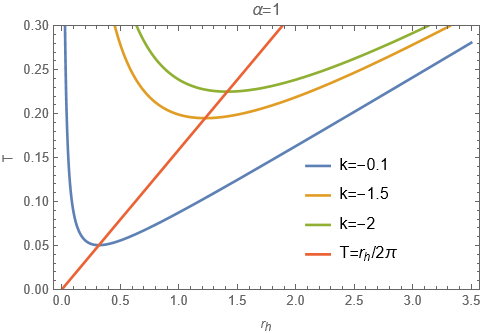}
    \caption{There are two branches of black hole solution and there is a positive minimum temperature corresponding to a negative k. Point set $\left\{r_h,T_{min}\right\}$ lies on the curve $T=\frac{r_h}{2 \pi}$ when $\alpha=1$.}
    \label{fig:2}
\end{figure}

In another case $1/\sqrt{3} < \alpha < \sqrt{3}$, as shown in Fig.\ref{fig:2}, the temperature can never reach zero. There is a minimum temperature at
\begin{equation}
    r_h=\sqrt{-k} \left(\frac{3 \alpha ^2-1}{3-\alpha ^2}\right)^{\frac{1-3 \alpha ^2}{2 \left(\alpha ^2+1\right)}}\,,
\end{equation}
and $T_{min}$ is
\begin{equation}
    T_{min}=\sqrt{-k}\frac{\sqrt{\left(3 \alpha ^2-1\right) \left(3-\alpha ^2\right)}}{2 \pi  \left(\alpha ^2+1\right)}\,.
\end{equation}
This indicates that $T$ is a function proportional to $r_h$ or $\sqrt{-k}$ with a given $\alpha$. For a given temperature above $T_{min}$, there are two values of $r_h$.
\par

\section{Thermodynamics and Unified First Law}
\label{sec.4}
We study the parameter of mass first. In the situation without angular deformation, there is a hyperbolic asymptotically AdS boundary and the mass of black hole can be strictly calculated in Einstein-scalar system by holographic renormalization (see \cite{Ren:2019lgw} and \cite{Caldarelli:2017} for details). 
\begin{equation}
    M=-\frac{\Omega}{8\pi G}(\frac{1-\alpha^2}{1+\alpha^2}b)\,.
    \label{eq:m0}
\end{equation}
The entropy is
\begin{equation}
    S=\frac{\Omega}{4 G}U(r_h)=\frac{\Omega }{4 G}{(-k)}^{\frac{2 \alpha ^2}{3 \alpha ^2-1}} r_h^{\frac{2 \left(\alpha ^2-1\right)}{3 \alpha ^2-1}}\,,
    \label{eq:S}
\end{equation}
where $\Omega$ is the size of the unit surface of angular space $d\Sigma^{2}_{2,k}$. The fisrt law of thermodynamics $dM-TdS=0$ is easily satisfied by equation(\ref{eq:T}), (\ref{eq:m0}), and(\ref{eq:S}) with the relation of k and $r_h$ in this situation. 

Although $M$ is exactly the thermodynamics mass in the asymptotically $AdS_4$ spacetime, it is no longer the "mass" for our solution, which lacks a suitable asymptotic AdS structure due to $e(\rho)$ contributing a nonconstant term in the curvature. The properties of these black holes are different from those with horizon having constant curvature. A method for calculating quasi-local mass without relying on global parameters is proposed in Ref.\cite{Hendi:2010}. This coincides with the viewpoint of the unified first law which based on the quasilocal Misner-Sharp(MS) mass. Consider a D-dimensional manifold $\overline{\mathcal{M}}_2 \times \hat{\mathcal{M}}_{D-2}$. Its line element is
\begin{equation}
\begin{split}
    ds^2&=g_{\mu \nu}dX^{\mu}dX^{\nu}\\
        &=\overline{g}_{ab}du^{a}du^{b}+\tilde{r}^2(u)\hat{g}_{ij}dx^{i}dx^{j}\,,
\end{split}
\end{equation}
where $\{u^a\}$ labels the point in the 2-dimensional radial part $\overline{\mathcal{M}}_2$ and $\{x^i\}$ labels the point in the $D-2$-dimensional angular part $\hat{\mathcal{M}}_{D-2}$ while $\tilde{r}$ is the areal radius. The overline denotes radial part while the hat denotes angular part. Both manifolds have independent metric $\overline{g}_{ab}$, $\hat{g}_{ij}$ and independent curvature $\overline{R}_{ab}$, $\hat{R}_{ij}$. In above ansatz, the MS mass is shown as:
\begin{equation}
        M_{MS}(u,x)=\frac{(D-2)\Omega}{16\pi G}\tilde{r}^{D-3}(\frac{\hat{R}}{(D-2)(D-3)}-\bar{g}^{\tilde{r}\tilde{r}})\,,
\end{equation}
where $\Omega$ is an area integral
\begin{equation}
    \Omega=\int_{S} \frac{\rho}{\sqrt{1-k \rho^2-e(\rho)}}d\rho d\varphi\,,
\end{equation}
and $\bar{g}^{\tilde{r}\tilde{r}}=\bar{\nabla}_a\tilde{r}\bar{\nabla}^a\tilde{r}$.

In order to use these methods, the metric should have the form 
\begin{equation}
\begin{split}
        ds^2=&-f(\Tilde{r})dt^2+\frac{1}{f(\Tilde{r})(\frac{d\Tilde{r}}{dr})^2}{d\Tilde{r}}^2\\
             &+{\Tilde{r}}^2[(\frac{1}{1-k\rho^2-e(\rho)})d\rho^2+\rho^2d\varphi^2]\,,
\end{split}
\end{equation}
where 
\begin{equation}
\Tilde{r}(r)=U(r)^{\frac{1}{2}}=r(1-\frac{b}{r})^{\frac{\alpha^2}{1+\alpha^2}}\,,
\end{equation}
and its MS mass should be
\begin{equation}
    \begin{split}
        M_{MS} (\tilde{r},\rho)=\frac{\Omega}{8\pi G}\tilde{r}\bigg(\frac{\hat{R}}{2}-f(\tilde{r})\tilde{r}'^2\bigg)\,,
    \end{split}
\end{equation}
where $\hat{R}$ is the Ricci scalar of angular space given by (\ref{angular R}).

We propose to incorporate an axion charge term into the first law of thermodynamics while ensuring the validity of the equation. The first method is to apply quasi-local viewpoint and unified first law to reengineer the equation $dM-TdS=0$. We need to define the energy supply vector $\Psi^{a}$ and the work term $W$ first in our ansatz
\begin{equation}
\begin{split}
    \Psi^a&=(T^{ab}-\frac{1}{2}T_{cd}\bar{g}^{cd}\bar{g}^{ab})\bar{\nabla}_{b}\tilde{r} 
          =-\frac{1}{8\pi G}f(r)\frac{\tilde{r}'\tilde{r}''}{\tilde{r}}\bar{\nabla}_{a}\tilde{r},\\
         W&=-\frac{1}{2}T_{ab}\bar{g}^{ab}=-\frac{1}{8\pi G \tilde{r}}\left(\bar{\nabla}^{2}\tilde{r}-\frac{1}{\tilde{r}}(\frac{\hat{R}}{2}-\bar{g}^{\tilde{r}\tilde{r}})\right)\,,
\end{split}
\end{equation}
%\frac{D-2}{2r}I_{ab}\left(\bar{\nabla}^{2}r-\frac{D-3}{r}(\mathcal{K}-I^{rr})\right) %
where $T_{\mu \nu}={T^{(em)}}_{\mu \nu}+{T^{(\phi)}}_{\mu \nu }+{T^{(\psi^{i})} }_{\mu \nu} $ is the total energy-momentum tensor.
The equation of unified first law becomes
\begin{equation}
    (dM_{MS})_{a}=W(dV)_{a}+A \Psi_{a}\,,
    \label{ufl}
\end{equation}
where the "volume" $V=\Omega r^{D-1}/(D-1)$ and area $A=\Omega r^{D-2}$. If let them take values on the horizon, the unified first law implies the first law of black hole thermodynamics in terms of such quasi-local parameters. The unified first law yields
\begin{equation}
   2\pi \frac{\delta M_{MS}-W_H\delta V_H}{(\kappa_{geo})_{H}}=\frac{\delta A_{H}}{4 G}=\delta S\,,
    \label{MS mass and M0 guanxi}
\end{equation}
where we use subscript $H$ to denote quasi-local physical quantities at the horizon and the surface gravity is $\kappa_{geo}\equiv \frac{1}{2}\bar{\nabla}^2\tilde{r}$. This equation implies that we can study the connection between the quasi-local viewpoint with the global viewpoint.

Therefore, the second approach attempts to establish a connection between the MS mass and the $AdS_4$ mass.
We focus on the situation at horizon where $f(\Tilde{r})$ goes to zero:
\begin{equation}
      M_{MS} (\tilde{r}_h,\rho)=\frac{\Omega \tilde{r}_h}{8\pi G}[k+\frac{8\pi G\xi^2}{\rho^2}]\,.
\end{equation}
Motivated by the structure of Equation (\ref{MS mass and M0 guanxi}), we find that MS mass can be seen as a linear combination of the mass of the EMD-AdS black hole and another term contributed by axions. 
\begin{equation}
    M_{MS}=M_{0}+M_{\xi}\,.
    \label{eq:M fenjie}
\end{equation}
We denote the EMD term with subscript $0$ and axion term with subscript $\xi$.
\par

When there are axions in EMD-ADS spacetime, we see the first law as \cite{Caldarelli:2017}:
\begin{equation}
    dM_{MS}+\varpi d \Pi + PdV=dM_0\,,
    \label{eq:thermo all}
\end{equation}
with $T$ the temperature on the horizon, $S$ the entropy of the horizon, $\varpi$ the axionic potential, $\Pi$ the axionic strength, $P$ the pressure, and $V$ the spatial volume.
\par
From (\ref{eq:M fenjie}) and (\ref{eq:thermo all}), we have
\begin{equation}
    dM_{\xi}+\varpi d\Pi+P_{\xi}dV=0\,.
\end{equation}
It is straightforward to see that all the considered physical quantities have only angular contributions, where the pressure has been defined as the conjugate variable to the volume, so that 
\begin{equation}
    P_{\xi}=(\frac{\partial M_{\xi}}{\partial V})_\varpi=\frac{3 a^2}{U(r)\theta^2}\,.
\end{equation}
$\Pi$ is defined as
\begin{equation}
    \Pi=\frac{1}{L}\sqrt{\frac{1}{D-2}\sum(\partial\psi^{i})^2}=\frac{a}{2\theta U^\frac{1}{2}}\,.
\end{equation}
After inserting the relation (\ref{eq:b}), the conjugate variable of axionic charge is
\begin{equation}
    \varpi=-\frac{a  (-k)^{\frac{2 \alpha ^2}{3 \alpha ^2-1}} r^{\frac{2 \left(\alpha ^2-1\right)}{3 \alpha ^2-1}}}{\theta }\Omega\,.
\end{equation}

\section{Conclusion}
\label{sec.5}
%正文里提一句爱因斯坦流形
In this study, we investigated the existence of EMD-AdS black hole solutions with deformed horizons. The successful construction of these solutions stems from the inherent decoupling of radial and angular coordinates in the Einstein equations for static black hole configurations, a critical feature that persists even the angular space is not an Einstein manifold. Such a feature is supported by the non-trivial configurations of two axion fields. Remarkably, this deformation is achieved without requiring modifications to the fundamental EMD action.
%In this article, we explored the possibility of EMD-AdS black hole with the deformed horizon. Due to the independence of the radial part and angular part of Einstein equations for non-rotating black holes, our exploration has been successful, without even change the action of EMD black hole. The fusion we have discussed are not meaningless, especially with AdS-like black holes. These fusion solutions are very interesting as the holographic duality relation persist.
\par
Unlike maximally symmetric spacetimes, in this solution the parameter $k$ cannot be normalized through rescaling angular coordinates. We have shown how the value of $k$ affect the horizon temperature. Furthermore, we studied the thermodynamics of our black hole solution. Specially, due to the lack of precise defination of the mass, we applied the MS mass and unified first law to the first law of thermodynamics in a quasi-local manner. It gives a formal expression for the first law in term of global parameters. We left this issue for future work. In addition, it is still a challenging problem to calculate the thermal conductivity in such a geometry with a deformed horizon for developing further application in AdS/CFT. Another aspect worth studying is how to construct the corresponding solution with angular momentum.

\section*{Acknowledgements}
Yusheng would like to thank %Jinbo Yang%
Hongbao Zhang, Zhan-Feng Mai, Jie Ren, Carlos A.R. Herdeiro for their helpful discussions or comments. This work is partially supported by Guangdong Major Project of Basic and Applied Basic Research (No.2020B0301030008). J. Yang is supported by the Guangzhou Postdoctoral Start-up Fund. 
%(致谢基金)%

%% The Appendices part is started with the command \appendix;
%% appendix sections are then done as normal sections
\appendix

%\section{Specific Cases of Similar Scalar Potentials}
%\setcounter{equation}{0}
%\renewcommand{\theequation}{A.\arabic{equation}}
%Among all similar or equivalent EMD-AdS solutions, the solution proposed by Anabalon and Astefanesei is the earliest we have identified. Their ansatz is as follows:
%(附录废案）

%% If you have bibdatabase file and want bibtex to generate the
%% bibitems, please use
%%
\bibliographystyle{unsrt}
\bibliography{mai}

%% else use the following coding to input the bibitems directly in the
%% TeX file.

%%\begin{thebibliography}{00}

%% \bibitem[Author(year)]{label}
%% For example:

%% \bibitem[Aladro et al.(2015)]{Aladro15} Aladro, R., Martín, S., Riquelme, D., et al. 2015, \aas, 579, A101

%%\end{thebibliography}

\end{document}